\documentclass[twocolumn,prb,superscriptaddress,preprintnumbers,showpacs]{revtex4}
\usepackage[latin9]{inputenc}
\usepackage{bm}
\usepackage{amsmath}
\usepackage{amssymb}
\usepackage{graphicx}

\makeatletter
\@ifundefined{textcolor}{}
{%
 \definecolor{BLACK}{gray}{0}
 \definecolor{WHITE}{gray}{1}
 \definecolor{RED}{rgb}{1,0,0}
 \definecolor{GREEN}{rgb}{0,1,0}
 \definecolor{BLUE}{rgb}{0,0,1}
 \definecolor{CYAN}{cmyk}{1,0,0,0}
 \definecolor{MAGENTA}{cmyk}{0,1,0,0}
 \definecolor{YELLOW}{cmyk}{0,0,1,0}
}

\usepackage{amsthm}\usepackage{amsfonts}\usepackage{algorithmic}\usepackage{enumerate}\usepackage{latexsym}\usepackage{bm}\usepackage{subfigure}\usepackage{ifthen}\usepackage{sidecap}

\makeatother

\begin{document}

\title{Algebraic and Geometric Mean Density of States in Topological Anderson
Insulators}

\author{Yan-Yang Zhang}

\affiliation{Department of Physics, The University of Hong Kong, Pokfulam, Hong
Kong, China}

\affiliation{SKLSM, Institute of Semiconductors, Chinese Academy of Sciences,
P.O. Box 912, Beijing 100083, China}

\author{Shun-Qing Shen}

\affiliation{Department of Physics, The University of Hong Kong, Pokfulam, Hong
Kong, China}

\date{\today}
\begin{abstract}
Algebraic and geometric mean density of states in disordered systems
may reveal properties of electronic localization. In order to understand
the topological phases with disorder in two dimensions, we present
the calculated density of states for disordered Bernevig-Hughes-Zhang
model. The topological phase is characterized by a perfectly quantized
conducting plateau, carried by helical edge states, in a two-terminal
setup. In the presence of disorder, the bulk of the topological phase
is either a band insulator or an Anderson insulator. Both of them
can protect edge states from backscattering. The topological phases
are explicitly distinguished as topological band insulator or topological
Anderson insulator from the ratio of the algebraic mean density of states
to the geometric mean density of states. The calculation reveals that
topological Anderson insulator can be induced by disorders from either
a topologically trivial band insulator or a topologically nontrivial
band insulator.
\end{abstract}


\maketitle

\section{Introduction}

Two-dimensional (2D) time-reversal invariant insulators can be classified
by $Z_{2}$ topological invariants for occupied energy bands \cite{Hasan2010,XLQiRMP,SQShen}.
Around the boundary of a system, a nontrivial topological invariant
guarantees the existence of gapless edge states lying in the bulk
gap between conduction and valence bands. The edge states are responsible
for the dissipationless transport of electrons, which persists even
in the presence of weak disorder. This is the picture of quantum spin
Hall effect or 2D topological band insulator (TBI) \cite{Kane-05prl,Bernevig2006,Zhou-08prl}.
Experimental observation of quantum spin Hall effect was reported
in recent years\cite{Konig2007,Roth2009,Knez-11prl,Du-13xxx}. On
the other hand, it has been further shown that the presence of disorder
makes the above picture more complicated: disorder can induce a topological
nontrivial phase from a trivial one, named as the topological Anderson
insulator (TAI)\cite{Li09,Groth2009,Jiang2009,HMGuo2010,Prodan2010,Prodan2011,YXXing2011,AYamakage2011,SRyu2012,DWXu,JTSong2012,YYZhang2012}.
As in conventional topological insulators, the dissipationless conductance
of TAI is also attributed to helical edge states. Nevertheless, there
is a remarkable difference: the robust edge states in TBI are protected
by an energy band gap while those in TAI are protected by a mobility
gap\cite{YYZhang2012,SQShen}. In other words, the TAI is a topological
state living in an Anderson insulator (AI), instead of a band insulator
(BI).

In this paper, we calculated the arithmetic and geometric mean density
of states (DOS) for a two-dimensional disordered topological insulator.
The DOS of a system describes the number of states per interval of
energy at each energy level that are available to be occupied by electrons.
Usually it is an average over the space and time. A zero value of
DOS at a finite interval of energy means that no states can be occupied
at that energy interval, which defines an energy gap. Consider the
local density of states (LDOS) $\rho(i)$ at each site $i$ of the
disordered sample. There are two types of mean DOS over sites. The
arithmetic mean of the LDOS over the sites of the sample
\begin{equation}
\rho_{\mathrm{ave}}\equiv\ll\rho(i)\gg\equiv\frac{1}{N}\sum_{i\in\mathrm{sample}}\rho(i)\label{eqrhoAve}
\end{equation}
is just the bulk DOS of the sample where $N$ is the total number
of sites. The geometric mean
\begin{equation}
\rho_{\mathrm{typ}}\equiv\exp[\ll\ln\rho(i)\gg]\label{eqrhoTyp}
\end{equation}
gives the typical value of the site LDOS over the sample. When electron
states are extended over the space, the DOS is almost uniform in the
space, and there is no distinct difference between the two means.
However, when electron states are localized, the local DOS is high
near some sites, but low near other sites. This fact will lead to
different mean values of DOS. While the DOS $\rho_{\mathrm{ave}}$
provides the information of energy levels, the ratio $\rho_{\mathrm{typ}}/\rho_{\mathrm{ave}}$
reveals the information of localization of the eigenstates: In the
thermodynamic limit ($N\rightarrow\infty$), if $\rho_{\mathrm{typ}}/\rho_{\mathrm{ave}}(E)\rightarrow$
finite, the states around $E$ is extended; while if $\rho_{\mathrm{typ}}/\rho_{\mathrm{ave}}(E)\rightarrow0$
, the states around $E$ is localized \cite{Weisse2006}.

Although both two-dimensional disordered TBI and TAI are characterized
by quantized conductance plateau in the measurement of two-probe setup,
we explicitly distinguish TAI from disordered TBIs from the arithmetic
mean DOS $\rho_{ave}$: there is an energy gap ($\rho_{\mathrm{ave}}$)
for the disordered TBI while $\rho_{ave}$ is finite for TAI. Starting
from either a normal band or an inverted band in the clean limit and
tuning disorder on gradually, we find that the TAI plays important
roles on the route towards final localization. Tuning on disorder
from a normal band, a disorder-induced band inversion will trigger
a topological Anderson insulating phase in which the bulk electrons
are localized. On the other hand, starting from an inverted band,
the topological phase can be divided into two well defined parts:
TBI and TAI, at weak and medium disorder strength, respectively.

\section{Model and General Formalism}

We start with the Bernevig-Hughes-Zhang model \cite{Bernevig2006}
on a square lattice, which is equivalent to the two-dimensional modified
Dirac equation \cite{Shen-11spin}. The model was proposed to describe
the quantum spin Hall effect in HgTe/CdTe quantum well, and has been
studied extensively. In the $k$ representation on a lattice with
periodic boundary condition, the Bloch Hamiltonian $\mathcal{H}$
is written as
\begin{eqnarray}
\mathcal{H}_{\alpha\beta}(\bm{k}) & = & \left(\begin{array}{cc}
h(\bm{k}) & 0\\
0 & h^{\ast}(-\bm{k})
\end{array}\right)_{\alpha\beta}\label{eqH}\\
h(\bm{k}) & = & d_{0}I_{2\times2}+d_{1}\sigma_{x}+d_{2}\sigma_{y}+d_{3}\sigma_{z}\notag\\
d_{0}(\bm{k}) & = & -2D\big(2-\cos k_{x}a-\cos k_{y}a\big)\notag\\
d_{1}(\bm{k}) & = & A\sin k_{x}a,\quad d_{2}(\bm{k})=-A\sin k_{y}a\notag\\
d_{3}(\bm{k}) & = & M-2B\big(2-\cos k_{x}a-\cos k_{y}a\big).\notag
\end{eqnarray}
Here $\alpha,\beta$ are the indices of electron spinorbitals within
the unit cell, $\alpha,\beta\in\{1,2,3,4\}\equiv\{|s\uparrow\rangle,|p\uparrow\rangle,|s\downarrow\rangle,|p\downarrow\rangle\}$.
$\sigma_{i}$ are the Pauli matrices acting on the spinor space spanned
by $s$ and $p$ orbitals. The real space Hamiltonian $H$ of this
model can be obtained from $\mathcal{H}_{\alpha\beta}(\bm{k})$ by
a straightforward inverse Fourier transformation. Throughout this
manuscript, we adopt the following model parameters from the HgTe/CdTe
quantum wells: $A=73.0$ meV nm, $B=-27.4$ meV, $C=0$, $D=-20.5$
meV, and lattice constant $a=5$nm\cite{Konig2008,Li09}. Then, the
parameter $M$ is a function of the thickness of quantum well: a positive
$M$ accounts for a topologically trivial phase while a negative $M$
for a topologically nontrivial phase in the clean limit as we take
a negative value for $B$ \cite{Lu-10PRB}. The effect of non-magnetic
impurities is included by introducing a term
\begin{equation}
V_{I}=\sum_{i}\sum_{\alpha}WU_{i}c_{i\alpha}^{\dagger}c_{i\alpha},\label{eqImpurity}
\end{equation}
where $U_{i}$ are random numbers uniformly distributed in $(-0.5,0.5)$
and $W$ is a single parameter to control the disorder strength. In
the following, we will call a definite realization of $U_{i}$ as
\emph{one} definite disorder sample.

The two-terminal conductance of the sample in units of $\frac{e^{2}}{h}$
is calculated by the standard non-equilibrium Green's function method\cite{Datta}
\begin{equation}
G_{L}(E_{F})=\text{Tr}(\Gamma_{L}G^{R}\Gamma_{R}G^{A}),\label{eqG}
\end{equation}
where the retarded (advanced) Green's function $G^{R(A)}(E_{F})=\big(E_{F}-H-\Sigma_{L}^{R(A)}-\Sigma_{R}^{R(A)}\big)^{-1}$,
$H$ is the Hamiltonian of the disordered sample, $\Gamma_{p}=i\big[\Sigma_{p}^{R}-\Sigma_{p}^{A}\big]$,
and $\Sigma_{p}^{R(A)}$ is the retarded (advanced) self energy from
lead $p(=L,R)$, i.e., the left and right lead. The topological phase
of the present quantum spin Hall model can be identified by the quantized
conductance $G_{L}=2\frac{e^{2}}{h}$.

The single particle local density of states (LDOS) of an isolated
sample at its site $i$ can be calculated explicitly by diagonalizing
the Hamiltonian of the disordered sample. After obtaining the energy
eigenstates for a specific disordered sample, $\left|n\right\rangle $,
where $n$ is the index for energy level, the LDOS is defined as
\begin{equation}
\rho(i,E_{F})=\sum_{n}|\langle i|n\rangle|^{2}\delta(E_{F}-E_{n}).\label{eqDOS}
\end{equation}
In numerical calculations, the Dirac delta function is approximated
as $\delta(E)=\eta/(\pi(E^{2}+\eta^{2}))$, with a small, but finite
broadening $\eta=+0$. With the LDOS in hand, both the arithmetic
mean $\rho_{\mathrm{ave}}$ and geometric mean $\rho_{\mathrm{typ}}$
can be calculated as defined in Eqs. (\ref{eqrhoAve}) and (\ref{eqrhoTyp}).

\section{Numerical Results}

\subsection{A normal band}

\begin{figure*}[htbp]
\fbox{\includegraphics[bb=28 37 870 665,clip,width=0.3\textwidth]{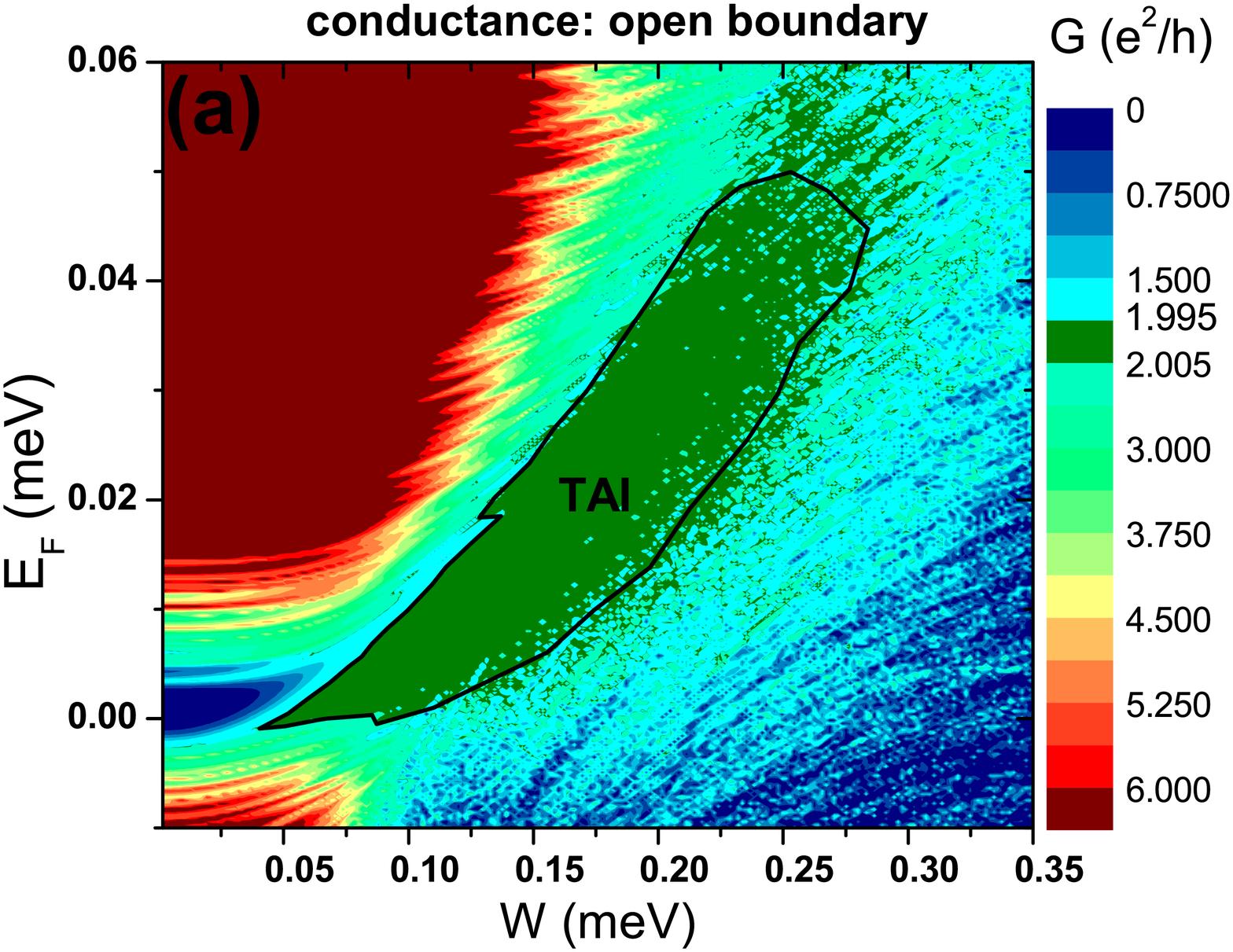}}
\fbox{\includegraphics[bb=28 37 870 665,clip,width=0.3\textwidth]{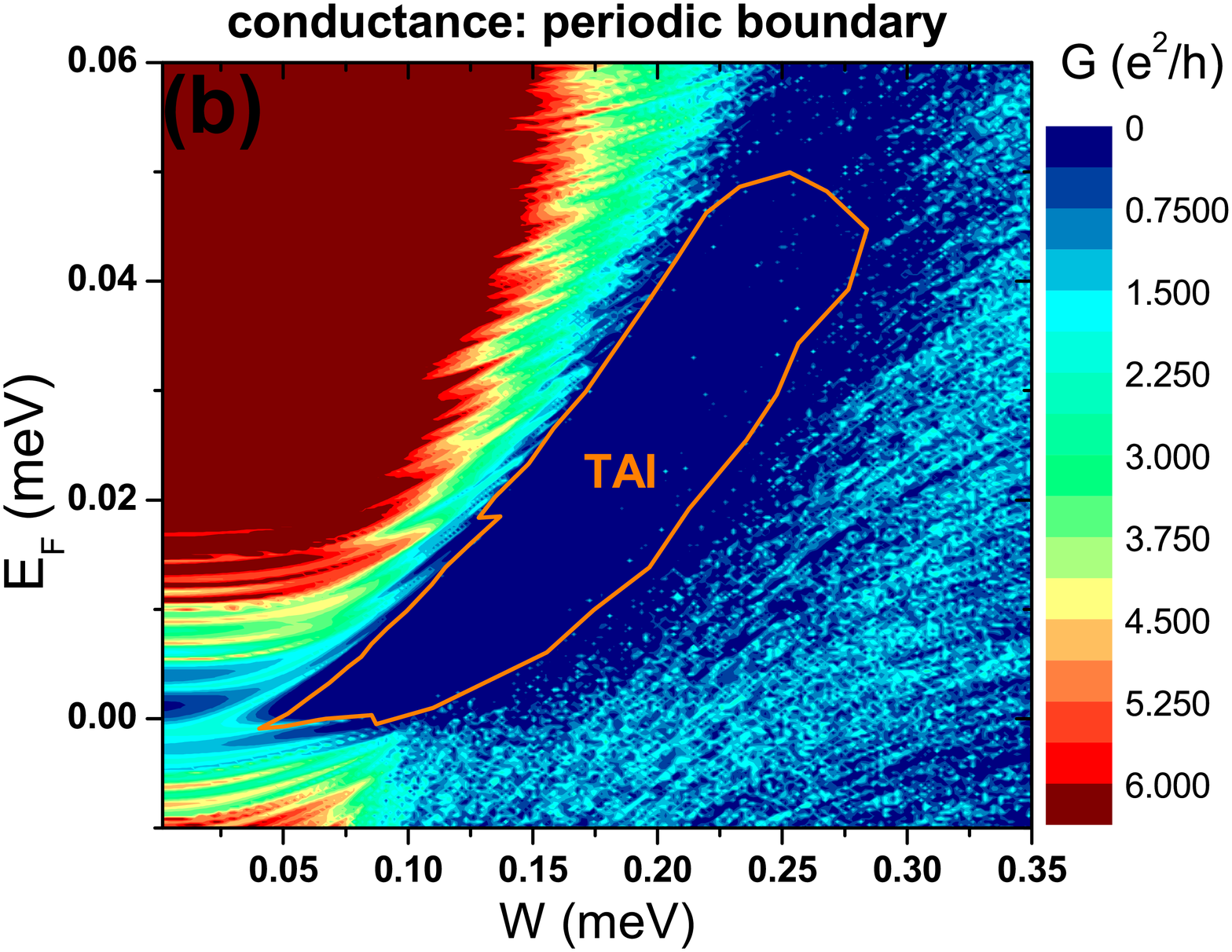}}\\
 \fbox{\includegraphics[bb=28 37 870 665,clip,width=0.3\textwidth]{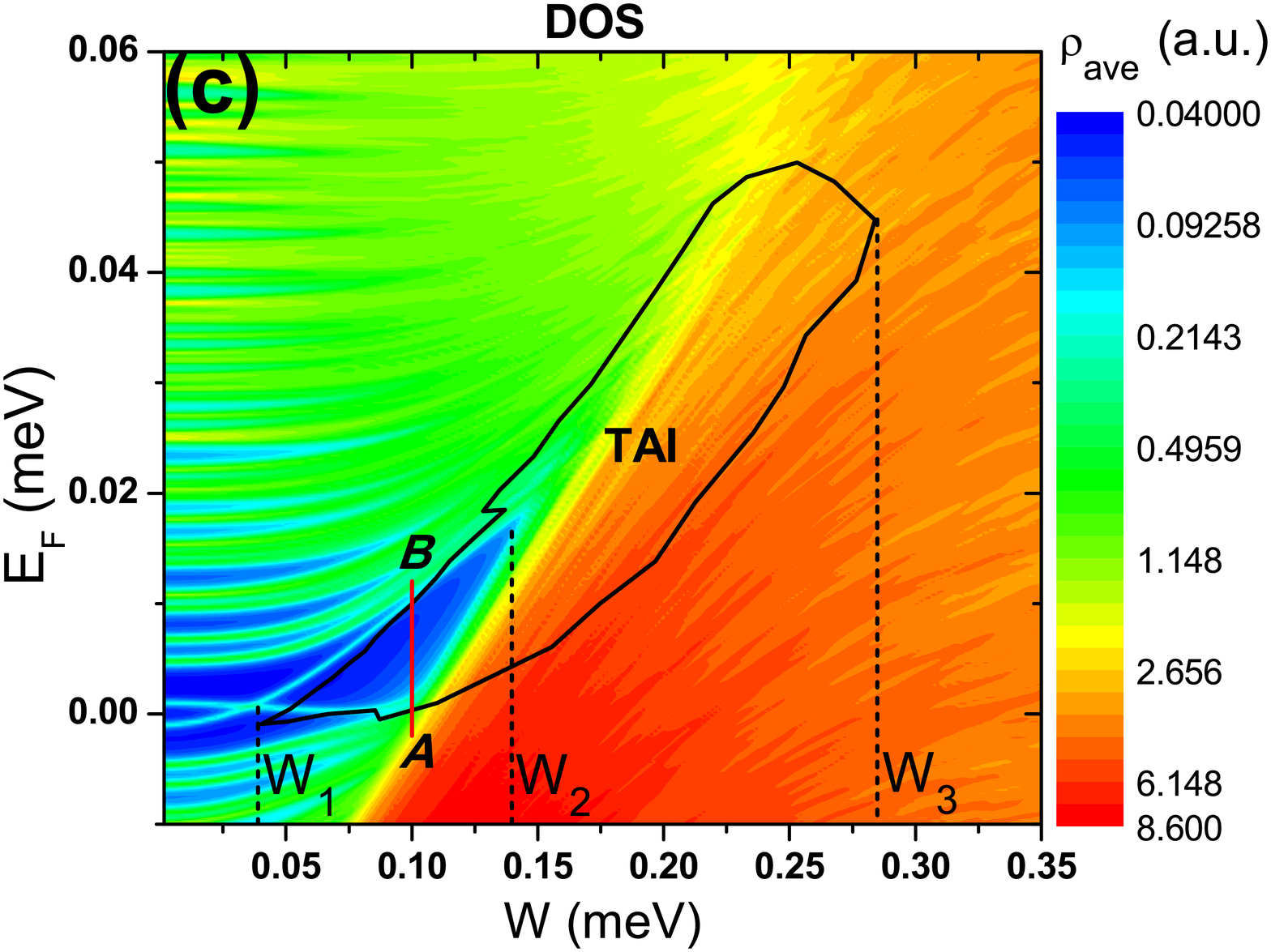}}
\fbox{\includegraphics[bb=28 37 870 665,clip,width=0.3\textwidth]{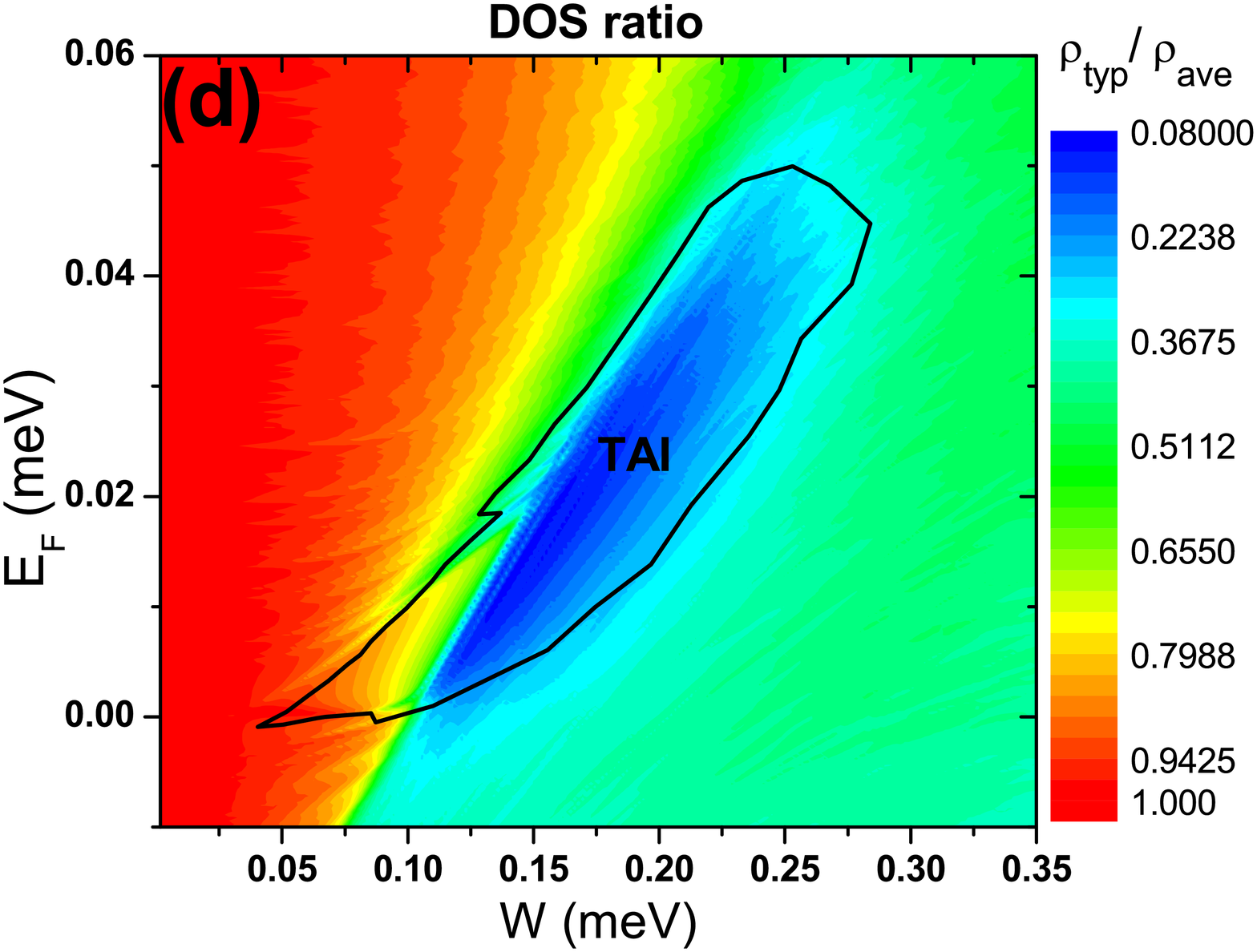}}
\caption{(Color online) Phase diagram in the parameter space of disorder strength
$W$ and Fermi energy $E_{F}$, for a normal band with $M=1$meV in
the clean limit. (a) Two-terminal conductance of a stripe sample with
an open boundary condition in the transverse direction; (b) Two-terminal
conductance of a rolled sample with the periodic boundary condition;
(c) the arithmetic mean DOS $\rho_{\mathrm{ave}}$; (d) the ratio
of the geometric mean DOS to arithmetic mean DOS: $\rho_{\mathrm{typ}}/\rho_{\mathrm{ave}}$.
Data in (c) and (d) are calculated numerically from the isolated sample
with the periodic boundary conditions in both directions, and with
energy broadening $\eta=2\times10^{-4}$meV. Other model parameters
are fixed throughout this work: $A=73.0$ meV, $B=-27.4$ meV, $C=0$,
$D=-20.5$ meV (see text). The sample size is of $100\times100$ sites,
and the lattice constant $a=5$nm. The boundary of the topological
phase in (c) and (d) is adapted from (a).}

\label{FigTAI} 
\end{figure*}

Let us start from the clean system with $M=1$ meV, which is topologically
trivial. When disorder strength $W$ is tuned on gradually, as first
found in Ref. \cite{Li09}, the system can evolve into TAI in a very
large range in the parameter space, which is characterized by perfectly
quantized conductance plateau $2e^{2}/h$ as expected to be carried
by the helical edge states. This disorder induced conductance plateau
is contrary to the general intuition that disorder always tends to
localize electronic states. Here we reconstruct the conductance diagram
on the $W-E_{F}$ plane for a definite sample with the open boundary
condition in Fig. \ref{FigTAI} (a). To compare with the bulk transport
mechanism, we also present the conductance diagram for the same sample
with the periodic boundary condition in the transverse direction in
Fig. \ref{FigTAI} (b). Notice the conductance $G_{L}=2\frac{e^{2}}{h}$
plateau in Fig. \ref{FigTAI} (a) corresponds to a bulk mobility gap
with $G_{L}\sim0$ in Fig. \ref{FigTAI} (b). This confirms once again
that the quantum transport in TAI is carried by helical edge states\cite{Li09,Jiang2009}.

In order to relate the above transport properties to the developments
of band structures of the sample, we present the arithmetic mean DOS
$\rho_{\mathrm{ave}}$ of the same sample in Fig. \ref{FigTAI} (c).
The developments of energy levels with the increasing of disorder
strength $W$ can be clearly illustrated. At small $W$, the conduction
and valance bands tend to get closer until they cross at $W_{1}\sim0.036$meV,
where a band inversion occurs. This gap closing and reopening indicates
a topological quantum phase transition induced by disorder\cite{Groth2009,YYZhang2012}.
This can be confirmed from Fig. \ref{FigTAI} (a) that after $W_{1}$,
there emerges the conductance plateau. Soon after the gap re-opening
$W_{1}$ in Fig. \ref{FigTAI} (c), one may notice a small inconsistency
between the angle-shaped boundary of conductance plateau (thick black
line) and the opened gap (blue region between the bands). This can
be simply attributed to different boundary conditions in the two calculations:
the DOS was calculated from an isolated sample while the conductance
was calculated in the presence of two leads connecting to the sample.
This boundary condition sensitivity in weak disorder regime can also
be observed in the same region Fig. \ref{FigTAI} (b).

\begin{figure}[htbp]
\includegraphics[bb=28 37 870 665,clip,width=0.4\textwidth]{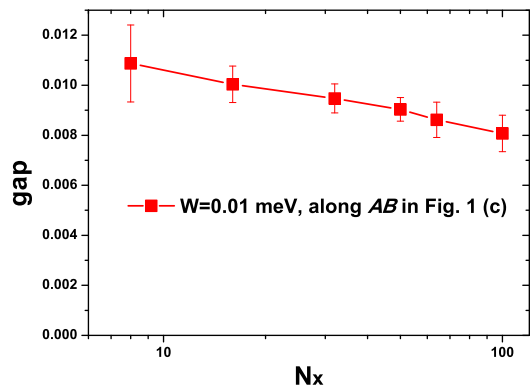}
\caption{(Color online) The energy gap at $W=0.1$meV (along the red line \emph{AB}
in Fig. \ref{FigTAI} (c)) as a function of lattice size $N_{x}$
for square samples. The $N_{x}$ axis is in logarithmic scale. Each
dot is a statistical average over 100 disorder samples.}

\label{FigTAIGap}
\end{figure}

However, by comparing Fig. \ref{FigTAI} (c) with the $G_{L}=2\frac{e^{2}}{h}$
plateau region (enclosed by the thick black line) obtained from Fig.
\ref{FigTAI} (a), we can see that this disorder induced band inversion
is not the whole story of the TAI: there is a level merging at $W_{2}\sim0.14$meV
after which the energy levels merge together without another gap reopening
again. This level merging does not affect the quantized conductance
at all. At a finite size (for example $N_{x}=N_{y}=100$ here), the
full gap regime ($\rho_{\mathrm{ave}}\sim0$) between $W_{1}$ to
$W_{2}$ (the blue region between $W_{1}$ and $W_{2}$ in Fig. \ref{FigTAI}
(c)) only constitutes a very small portion of the conductance plateau.
To investigate the fate of this band gapped topological insulator
regime, i.e. TBI, under the thermodynamic limit ($N_{x},N_{y}\rightarrow\infty$),
we plot the size dependence of the averaged gap size at $W=0.1$meV
(along line \emph{AB} in Fig. \ref{FigTAI} (c)) in Fig. \ref{FigTAIGap}.
This curve is approximately linear with $N_{x}$ in logarithmic scale,
therefore the gap decays slowly (logarithmically)
with the increasing of sample size $N_{x}$, suggesting
that this small ``TBI'' region shrinks smaller or even vanishes
in the thermodynamic limit.

Most part of the conductance plateau in Fig. \ref{FigTAI} corresponds
to a region with a large bulk DOS $\rho_{ave}$. In other words, these
topological states do not live in a bulk gap at all. From the plot
of ratio $\rho_{\mathrm{typ}}/\rho_{\mathrm{ave}}$ in Fig. \ref{FigTAI}
(d), we can see that this region of TAI is related to extremely localized
states with $\rho_{\mathrm{typ}}/\rho_{\mathrm{ave}}\rightarrow0$.
From Fig. \ref{FigTAI} (d), it is interesting to note that $\rho_{\mathrm{typ}}/\rho_{\mathrm{ave}}$
in the TAI region is even smaller (i.e., more localized) than any
other region (even with larger $W$) in the plot range. This is consistent
with the physical picture found in our previous work\cite{YYZhang2012},
that these extremely localized states (Anderson insulator) correspond
to flat but densely distributed subbands (therefore contribute to
DOS, but not to conductance). These flat subbands have trivial topological
numbers, therefore they do not change the topological nature of nearby
energy region. With open boundaries, helical edge states appear between
adjacent flat subbands and contribute to the conductance $2e^{2}/h$.
These subbands are so flat that their total measure on the energy
axis is extraordinarily small, therefore there is always enough space
to occupy helical edge states between them, in the case of an open
boundary. In other words, the TAI phase lives in a bulk state of Anderson
insulator (instead of band insulator), as the name suggests. The extremely
localized nature of the bulk Anderson insulator also prohibits backscattering
between edge states on opposite edges, therefore we call that the TAI is
protected by a mobility gap instead of a band gap.

With further increasing of disorder strength $W$, as in any lattice
systems, the electronic states will finally be localized completely
and the system becomes an Anderson insulator \cite{Hatsugai1999},
which happens after $W_{3}\sim0.28$meV in the present system.

\begin{figure*}[htbp]
\fbox{\includegraphics[bb=28 37 870 665,clip,width=0.3\textwidth]{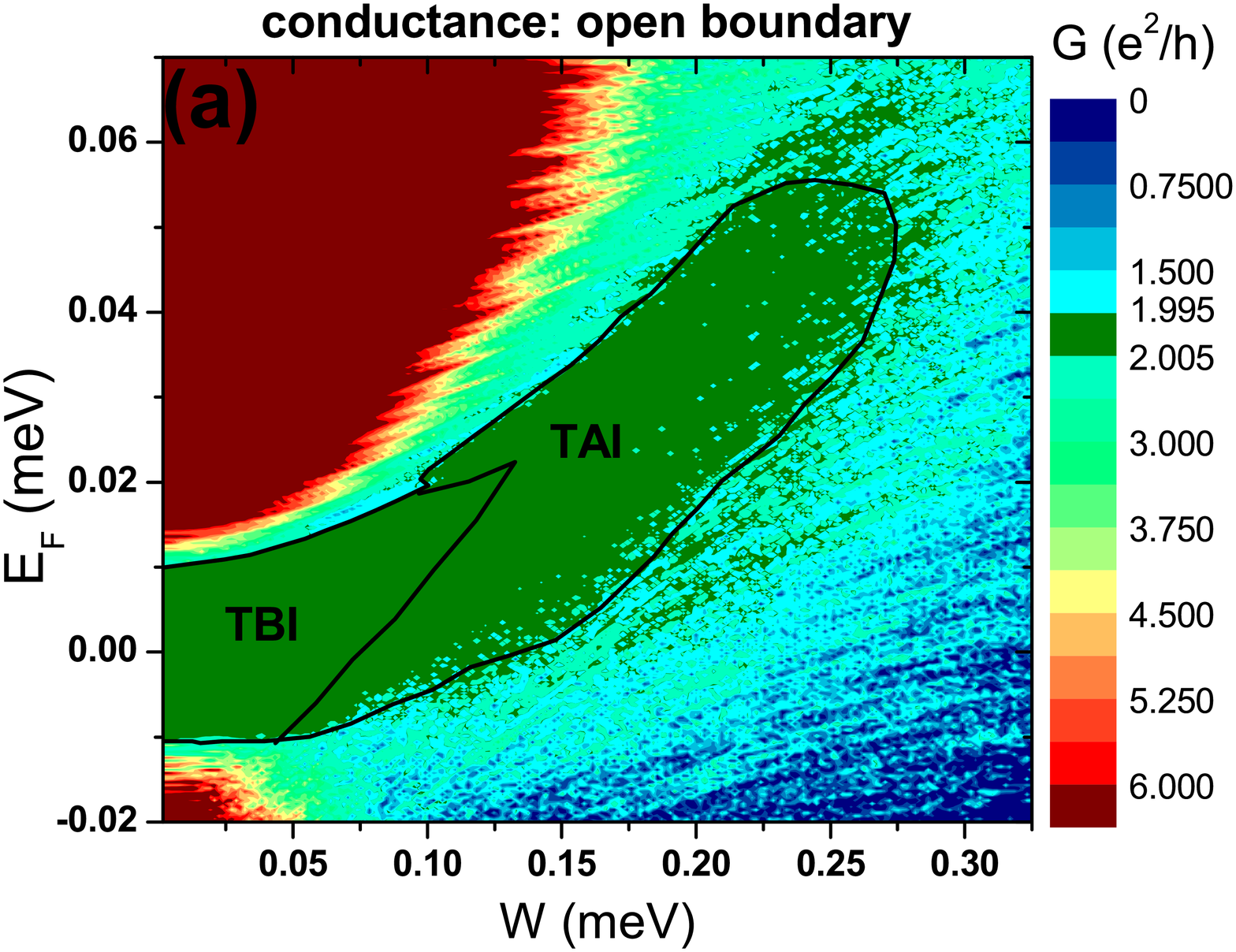}}
\fbox{\includegraphics[bb=28 37 870 665,clip,width=0.3\textwidth]{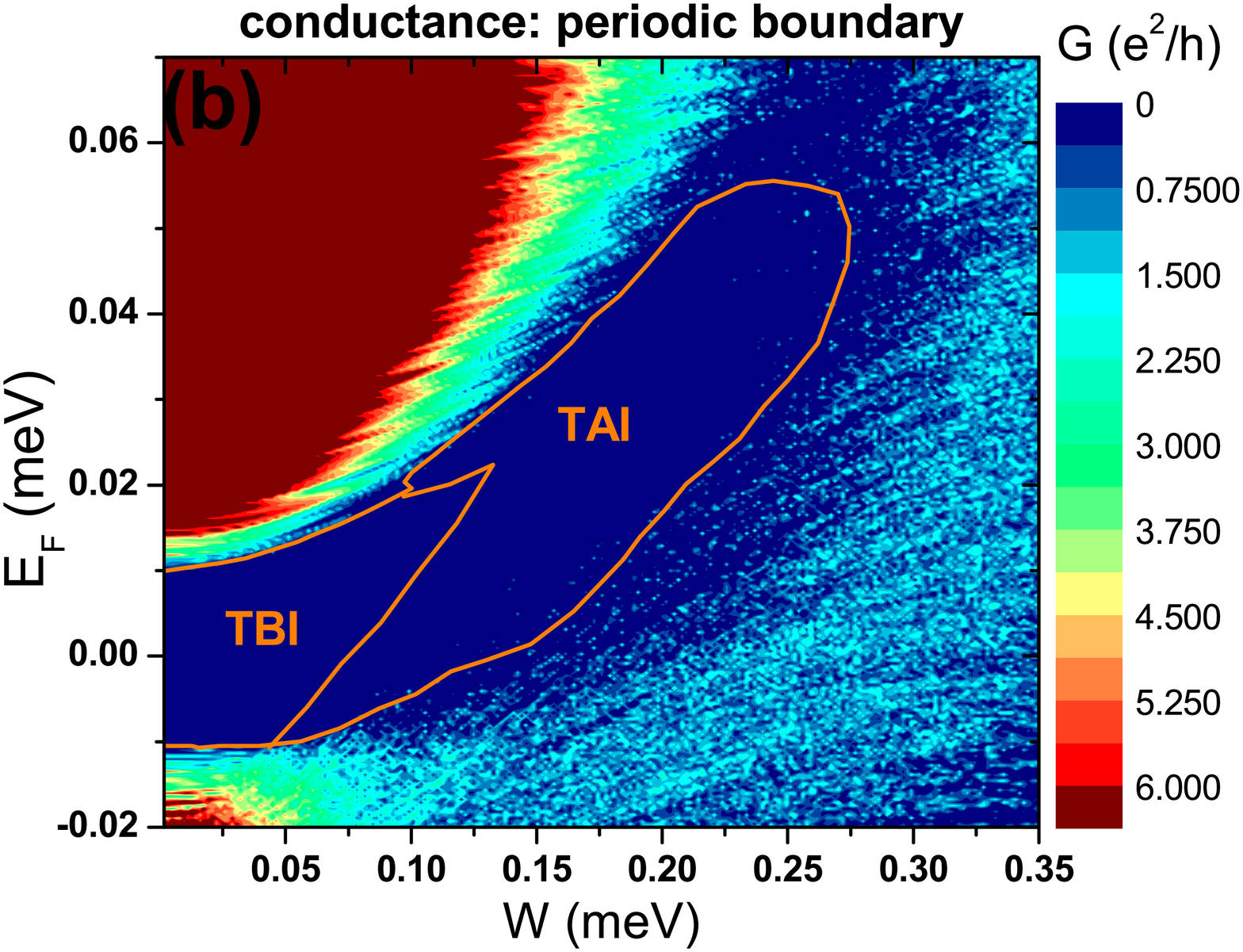}}\\
 \fbox{\includegraphics[bb=28 37 870 665,clip,width=0.3\textwidth]{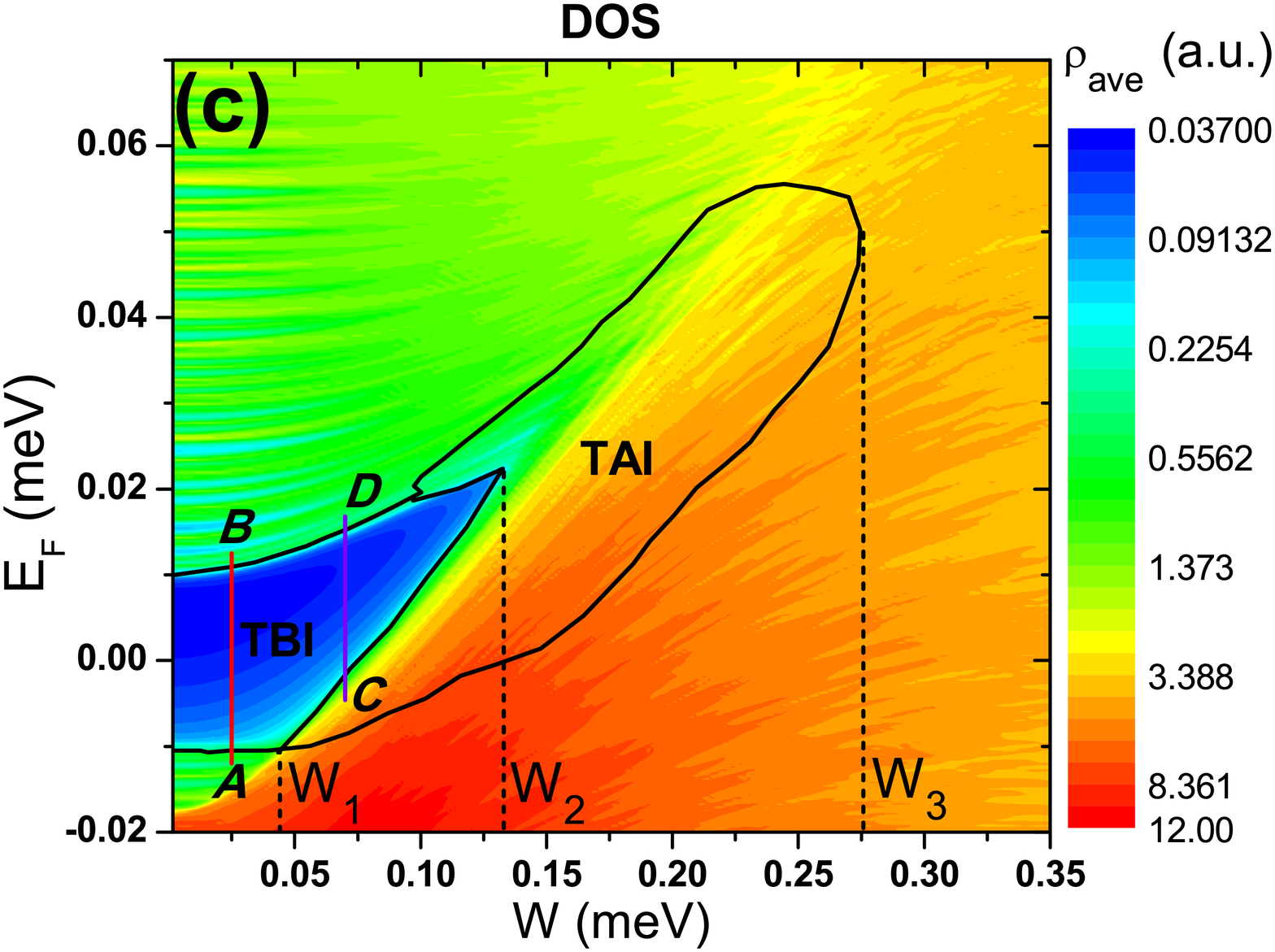}}
\fbox{\includegraphics[bb=28 37 870 665,clip,width=0.3\textwidth]{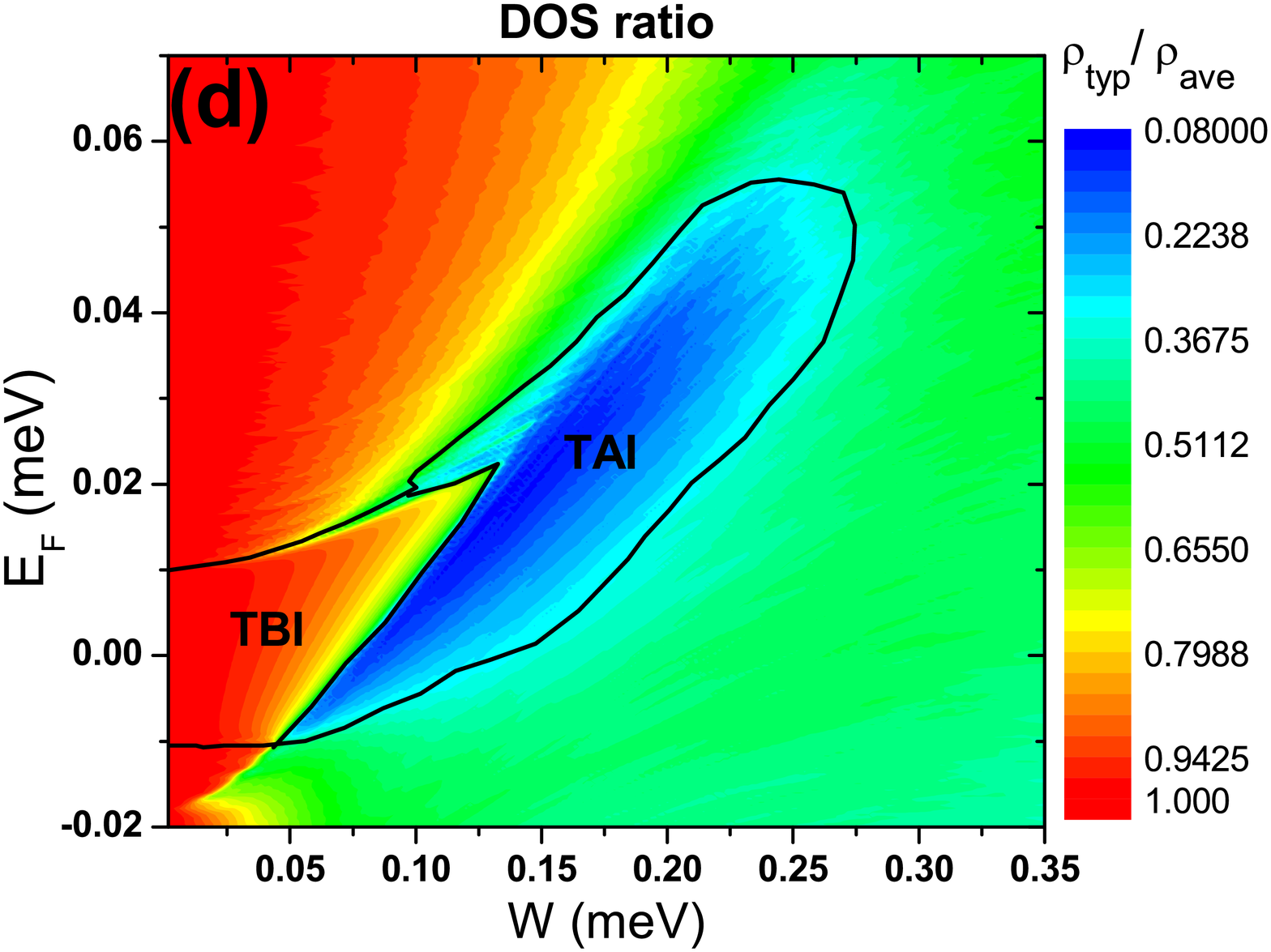}}
\caption{(Color online) Phase diagram in the parameter space of disorder strength
$W$ and Fermi energy $E_{F}$, for an inverted band with $M=-10$meV
in the clean limit. (a) Two-terminal conductance of a stripe sample
with an open boundary condition in the transverse direction; (b) Two-terminal
conductance of a rolled sample with the periodic boundary condition;
(c) the arithmetic mean DOS $\rho_{\mathrm{ave}}$; (d) the ratio
of the geometric mean DOS to arithmetic mean DOS: $\rho_{\mathrm{typ}}/\rho_{\mathrm{ave}}$. }

\label{FigTI}
\end{figure*}

\subsection{An inverted band}

Now we turn to the case of an inverted band:$M=-10$ meV, which is
a TBI in the clean limit. From the conductance plateau in Fig. \ref{FigTI}
(a) and (b), it seems that this phase diagram is quite simple: the
topological phase characterized by helical edge states survives robustly
until a complete collapse at a very large disorder $W_{3}\sim0.28$meV.
However, a detailed numerical calculation demonstrates that the regime
of topological phase actually also consists of two parts: one is
the TBI and the other is TAI. Although the conductance plateau in
Fig. \ref{FigTI} (a) is continuous, the DOS can be used to distinguish
the two phases very well. In the DOS plot in Fig. \ref{FigTI} (c),
we can see the developments of energy levels under increasing of disorder
strength $W$. Note the conductance and valence bands merge at $W_{2}\sim0.13$meV,
which is much weaker than $W_{3}$ where the conductance plateau collapses.

\begin{figure}[htbp]
\includegraphics[bb=28 37 875 665,clip,width=0.4\textwidth]{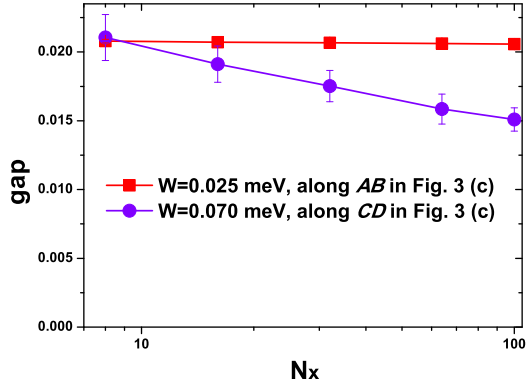}
\caption{(Color online) The energy gap at $W=0.1$meV (along lines \emph{AB}
(red square dot) and \emph{CD} (violet circle dot) in Fig. \ref{FigTI}
(c)) as a function of square samples with lattice size $N_{x}$. The
$N_{x}$ axis is in logarithmic scale. Each dot is a statistical average
over 100 disorder samples.}

\label{FigTIGap}
\end{figure}

We first concentrate on the conductance plateau region with $W<W_{2}$,
i.e., before the band merging, which is apparently a band insulator.
As can be seen in Fig. \ref{FigTI} (c) for a definite size and disorder
configuration, at weak disorder starting from $W=0$, the bulk gap
increases slowly with increasing disorder until $W_{1}$, reflecting
the band repelling of topological nontrivial bands under disorder\cite{YYZhang2012}.
On the other hand, as the red curve with square dots in Fig. \ref{FigTIGap},
the size dependence of the averaged gap at a definite $W$ (along
line \emph{AB} in Fig. \ref{FigTI} (c) ) indicates that it is a horizontal
line with negligible statistical errors, suggesting a robust region
of TBI independent of sample size and disorder configurations. After
$W_{1}$, the bulk gap begins to shrink, but the conductance plateau
remains unchanged. In other words, TAI with quantized $G_{L}=2\frac{e^{2}}{h}$
persists and finite $\rho_{\mathrm{ave}}$ gradually appears when
$W>W_{1}$. In Fig. \ref{FigTIGap}, the gap size in this region for
a definite $W$ (along line \emph{CD} in Fig. \ref{FigTI} (c) ) as
a function of sample size is plotted as the violet curve: it decays
logarithmically. This simply suggests that in the thermodynamic limit,
the region of the TBI is even smaller than that plotted in Fig. \ref{FigTI}
for finite size $N_{x}=N_{y}=100$. On the other hand, as discussed
above, the TBI is always robust and well-defined before $W_{1}$.

On the right side of the border between TBI and TAI, the phase possesses
the typical characters of TAI: quantized conductance plateau living
in a bulk mobility gap with nonzero DOS from extremely localized states.
From the phase diagram in Fig. \ref{FigTI}, It is interesting to
notice that TAI phase (with quantized $G_{L}=2\frac{e^{2}}{h}$, finite
$\rho_{\mathrm{ave}}$ and small $\rho_{\mathrm{typ}}/\rho_{\mathrm{ave}}$)
still occupies a large portion of the conductance plateau. This comes
from the localization of bulk sates near band edges under disorder.
Therefore we emphasize that even tuning disorder on from an inverted
band, the TAI also plays a very important role in the whole topological
phase. As can be seen from Fig. \ref{FigTI}, the boundary between
the two topological phases is sharp and clear without any intermediate
state, say, a metal phase\cite{Prodan2011}.

\section{Summary and Discussions}

\begin{figure}[htbp]
\includegraphics[clip,width=0.42\textwidth]{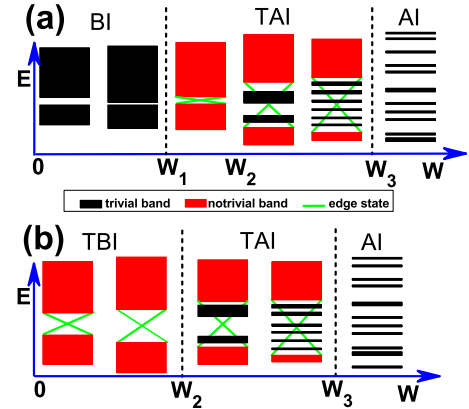} \caption{(Color online) Schematic of the evolution of band (energy level) structures
with disorder strength, starting from (a) a conventional band insulator;
(b) a topological band insulator. Topological trivial (nontrivial)
bands are represented by black (red) bars, and the green lines represents
the presence of the helical edge states between the two topologically
nontrivial bands.}

\label{FigSchematic}
\end{figure}

Now we can summarize the route towards localization schematically
fora disordered quantum spin Hall system in Fig. \ref{FigSchematic}.
In the clean limit of $W=0$, the electronic states form well-defined
energy bands, and the band gap between the conduction and valence
band are well defined. If we start from a conventional band insulator,
as shown in Fig. \ref{FigSchematic} (a), weak disorder tends to shrink
the bulk band gap\cite{Groth2009,YYZhang2012,DWXu}. If the gap is
small enough, then this disorder induced band attraction will eventually
lead to a band closing at $W_{1}$, after which edge states emerge
in the re-opened gap. On the other hand, disorder destroys the translation
invariance, making the energy band worse defined and split into more
discrete energy levels. Around the band edges, these levels constitute
to localized states\cite{Mott1968,Localization1993}. The calculation
of a single impurity indicates that the in-gap bound states can be
induced easily by a single impurity in the inverted band regime\cite{JLu2011}. After
$W_{2}$, these localized states permeate into the bulk gap, leading
to a well-defined TAI.

From this picture, we can see the condition for a stable disorder
induced TAI phase. First, the clean limit gap should be sufficiently
small to guarantee $W_{1}<W_{3}$. Second, after $W_{1}$, the in-gap
bulk states should be sufficiently localized (or equivalently, sufficiently
flat). From the viewpoint of momentum space, the in-gap states should
be flat enough for edge states between them to show up. From the view
point of real space, these states in the bulk should be localized
enough to protect helical edge states from backscattering. This is
why we call TAI a mobility gap protected topological insulating state,
compared with TBI, an energy gap protected one. When we start from
topological insulating phase, as in Fig. \ref{FigSchematic} (b),
weak disorder tend to expel the conduction and valance bands\cite{YYZhang2012}.
Before the disorder decompose the energy bands into sufficiently discrete
energy levels ($W<W_{2}$), the system is still in a well-defined
topological band insulating phase. After that, the bulk states near
the band edges will be localized\cite{Mott1968,Localization1993},
and grow into the bulk gap to make a TAI phase.

In this work, we try to distinguish the topological insulating phases
in 2D as a band insulator and an Anderson insulator. In Fig. 3 we
define a border between TBI and TAI in the inverted band structure.
This raises a question of whether they are physically distinct from
each other by measuring the mobility gap or any other physical quantities.
Usually temperature-dependencies of the bulk conductance are different
between band insulators and Anderson insulators. With the help of
thermal activation and phonons, the typical temperature dependence
of conductivity for the band insulator is $\sigma\sim\exp[-E_{g}/(2k_{B}T)]$
while that of the Anderson insulator is $\sigma\sim\exp[-(T_{0}/T)^{1/3}]$\cite{Mott1969}.

The energy scale of the two-dimensional mobility gap can be determined
from capacity measurement as well as transport measurement in Corbino
samples.\cite{Yang-97prl} In the geometry of Corbino disk, edge transport
is stuck around the concentric contacts, and the radial conductance
measurement mainly probe the bulk properties. The mobility gap can
be revealed by using capacitance-voltage and conductance measurement.
In recent experiment in InAs/GaSb bilayers \cite{Du-13xxx}, it is
observed that the existence of quantized conductance in quantum spin
Hall effect is accompanied by the appearance of the mobility gap.
The measured capacitance shows clearly the gap opening between conduction
and valance bands, and further opening of the mobility gap at lower
temperatures. Coexistence of the mobility gap and the quantized conductance
from helical edge states is characteristic of TAI. In this sense,
the quantized conductance appears in the TAI regime.
\begin{acknowledgments}
This work was supported by the Research Grant Council of Hong Kong
under Grant No.: HKU 7051/11P. YYZ was also supported by NSFC under
grant number: 11204294 and 973 Program Project No. 2013CB933304.\end{acknowledgments}

\end{document}